\documentclass[12pt]{article}
\usepackage{amsfonts}
\newlength{\abstractwidth}
\usepackage{epsf}

\renewcommand{\thefootnote}{\fnsymbol{footnote}}
\renewcommand{\thanks}[1]{\footnote{#1}}
\newcommand{\starttext}{
\setcounter{footnote}{0}
\renewcommand{\thefootnote}{\arabic{footnote}}}

\newcommand{\bea}{\begin{eqnarray}}
\newcommand{\eea}{\end{eqnarray}}
\newcommand{\ee}{\end{equation}}
\newcommand{\be}{\begin{equation}}

\def\half{ {1\over 2}}
\def\p{\partial}

\def\tet{\vartheta}

\def\ch{{\rm ch }}
\def\sh{{\rm sh }}

\def\tg{{\rm tg  }}

\def\no{\nonumber}
\def\sm{\smallskip}

\begin{document}
\starttext
\setcounter{footnote}{0}

\begin{flushright}
UCLA/08/TEP/23 \\
23 July 2008
\end{flushright}

\vskip 0.3in

\begin{center}

{\Large \bf Integrable systems from supergravity BPS 
equations}\footnote{This work was supported in part by National 
Science Foundation (NSF) grants PHY-04-56200\\
and PHY-07-57702.}

\vskip 0.7in

{\large Eric D'Hoker and John Estes}

\vskip .2in

 \sl Department of Physics and Astronomy \\
\sl University of California, Los Angeles, CA 90095, USA

\end{center}

\vskip .5in

\begin{abstract}

Integrable systems of the sine-Gordon/Liouville type, which arise from reducing
the BPS equations for solutions invariant under 16 supersymmetries in Type IIB 
supergravity and M-theory, are shown to be special cases of an infinite family 
of integrable systems, parametrized by an arbitrary real function $f$ of a real 
variable. It is shown that, for each function $f$, this generalized integrable 
system may be mapped onto a system of linear equations, which in turn may 
be integrated in terms of the two linearly independent solutions of an ordinary 
linear second order differential equation which depends only on the function $f$.

\end{abstract}

\newpage

\vfill\eject

\baselineskip=15pt
\setcounter{equation}{0}
\setcounter{footnote}{0}

\section{Introduction}
\setcounter{equation}{0}

A variety of BPS equations for solutions with 16 supersymmetries in Type IIB
supergravity in 10 dimensions and in M-theory in 11-dimensions are amenable
to general classification \cite{Gauntlett:2002sc,Gauntlett:2002,Gauntlett:2004zh},
and to exact solution \cite{Lin:2004nb,D'Hoker:2007xy,D'Hoker:2007fq,D'Hoker:2008wc}.
Exact solution could be achieved because the BPS equations can be mapped onto
integrable two-dimensional field theories of the sine-Gordon, Liouville or Toda type,
which in turn can be mapped onto linear problems. In this note, we shall show that
the various integrable systems of \cite{D'Hoker:2007xy,D'Hoker:2007fq,D'Hoker:2008wc} naturally fit into an infinite family of integrable  systems which are
all of the sine-Gordon/Liouville type. Each system is labeled uniquely by a real
function $f$ of a single real variable. For each $f$, we shall exhibit a local map of the
integrable system onto a linear system which may be integrated exactly
in terms of the two linearly independent solutions of a fixed ordinary linear second order
differential equation depending only on the function $f$.

\sm

The integrable systems resulting from the reduced BPS equations in
both M-theory and Type IIB supergravity are given in terms of a real field $\tet$,
and a real harmonic function $h$, which are both functions on a 2-dimensional
Riemann surface $\Sigma$ with boundary. For the M-theory problem, the field 
equation for $\tet$ is given  \cite{D'Hoker:2008wc} by,
\bea
\label{M}
{\rm M} ~~ & \hskip 0.6in & 4  \p_{\bar w} \p_w \tet
- \p_{\bar w} \bigg ( i e^{2 i \tet} \p_w \ln h \bigg )
+ \p_w \bigg (  i  e^{- 2i \tet} \p_{\bar w} \ln h \bigg )
 =0
\eea
For the Type IIB problem,\footnote{To exhibit more closely the
relation between  (\ref{M}) and (\ref{IIB}), the variables $\tet$ and $\mu$ used in
\cite{D'Hoker:2007xy} have been redefined by letting $\tet \to - 2 \tet + \pi/2$
and $\mu \to h$.}  the field equation for $\tet$ is given by \cite{D'Hoker:2007xy},
\bea
\label{IIB}
{\rm IIB} & \hskip 0.6in & 4  \p_{\bar w} \p_w \tet
- \p_{\bar w} \bigg ( 2i  { \p_w h \over \cos (h)} e^{2 i \tet}  \bigg )
+ \p_w \bigg ( 2i  { \p_{\bar w} h \over \cos (h)} e^{- 2i \tet}  \bigg )
 =0
\eea
Both systems may be viewed as special cases of an infinite family of
integrable systems of one real field $\tet$, labeled by a real function $f$
of one real variable, whose field equation is
\bea
\label{f}
4  \p_{\bar w} \p_w \tet
- \p_{\bar w} \bigg ( i e^{2 i \tet} \p_w \ln f(h) \bigg )
+ \p_w \bigg (  i  e^{- 2i \tet} \p_{\bar w} \ln f(h) \bigg )
 =0
\eea
Systems (\ref{M}) and (\ref{IIB}) respectively correspond to the functions
\bea
\label{MIIB}
{\rm M} &  \hskip 0.6in & f(h) = h
\no \\
{\rm IIB} && f(h) = \tg \left ( { h \over 2 } + {\pi \over 4} \right )^2
\eea

\section{General solution by mapping to a linear system}
\setcounter{equation}{0}

For any function $f$, the field equation (\ref{f}) is invariant under local 
conformal transformations of $w$ on $\Sigma$. The equation is of the 
sine-Gordon type in that it involves both types of exponentials $e^{ \pm 2 i \tet}$ (see for example \cite{SG1, SG2},
and references therein). In view of the explicit $h$-dependence, however, the system
is also akin to Liouville theory with unbroken time  translation invariance (along the
direction perpendicular to the coordinate direction $h$), but broken  space translation
invariance (parallel  to the coordinate direction $h$) \cite{Liouville1, Liouville2, Liouville3}. 
In particular, the system shares with Liouville theory the remarkable property that it may be
solved exactly (see \cite{Liouville1} and references therein), in a manner that will be 
made clear below.

\sm

Integrability of (\ref{f}) for any real function $f$ is guaranteed by the fact that
the field equation for $\tet$ may be recast in the form of a flatness condition,
\bea
\label{int}
\p_{\bar w} \bigg ( 2 \p_w \tet -  i e^{2 i \tet} \p_w \ln f(h) \bigg )
+
\p_w \bigg (  2 \p_{\bar w} \tet + i  e^{- 2i \tet} \p_{\bar w} \ln f(h) \bigg ) =0
\eea
As a result, there must exist a  real function $\psi$, which is locally defined
on $\Sigma $, such that
\bea
\label{bak}
 2 \p_w \tet   - i e^{2 i \tet} \p_w \ln f(h)  & = & + 2 i \p_w \psi
\no \\
 2 \p_{\bar w} \tet  +i   e^{- 2i \tet} \p_{\bar w} \ln f(h)  & = & - 2 i \p_{\bar w} \psi
\eea
Integrability of (\ref{bak}), viewed as a system of differential equations for $\psi$, gives
the field equations (\ref{f}) or equivalently (\ref{int}). Integrability in $\tet$
yields an algebraic relation between $\tet$, $f$ and $\p_w \p_{\bar w} \psi$,
which will not be used here.

\sm

To integrate the system of first order equations (\ref{bak}),
the methods of \cite{D'Hoker:2008wc}
may be generalized to the problem at hand. A simple change of variables allows
one to map (\ref{bak}) onto a linear problem for any function $f$. We introduce
\bea
F \equiv \sqrt{f} \, e^{- \psi - i \tet}
\eea
and its complex conjugate $\bar F$. In terms of $F$, the system (\ref{bak}) of
non-linear first order equations is mapped into a system of linear equations,
\bea
\label{F}
\p_w F = \half (F + \bar F) \, \p_w \ln f(h)
\eea
and its complex conjugate. This equation coincides with equation (5.41) of
\cite{D'Hoker:2008wc}, except for the dependence on $h$, which is now through
a general function $f(h)$. Even with this extra $f$-dependence, (\ref{F})
is manifestly linear, conformal invariant, and may be solved as follows.

\sm

Using conformal reparametrization invariance of (\ref{F}) and the fact
that $h$ is real and harmonic, we may choose adapted coordinates
$w = r + i x$ where $r=h$, and $x=\tilde h$ is  the harmonic
function dual to $h$, satisfying $\p_{\bar w} (h+ i \tilde h)=0$. Decomposing
the complex function $F=F_r + i F_x$ into its real and imaginary parts $F_r$
and $F_x$, (\ref{F}) decomposes as follows,
\bea
\label{pF}
\p_r F_x - \p_x F_r & = & 0
\no \\
\p_r F_r + \p_x F_x & = & { f'(r) \over  f(r)} \,  F_r
\eea
The first equation in (\ref{pF}) may be solved in terms of a single real
function $\Phi(r,x)$, and we have $F_r = \p_r \Phi$ and $F_x = \p_x \Phi$.
In terms of $\Phi$, the second equation becomes,
\bea
\left ( \p_r^2 + \p_x ^2 - { f'(r) \over f(r) } \p_r \right ) \Phi (r,x)=0
\eea
Thanks to translation invariance in $x$, we may solve for the $x$-dependence of $\Phi$
in terms of a Fourier transform in this variable. Using also the reality of the function $\Phi$,
the Fourier transform takes the following form,
\bea
\Phi (r,x) =
\int _0 ^\infty { dk \over 2 \pi} \bigg  ( \Phi _k (r) e^{-ikx} + \Phi ^* _k (r) e^{+ikx} \bigg )
\eea
The Fourier modes $\Phi _k (r)$ are complex-vlaued functions of $r$ and $k$,
and must satisfy the following {\sl ordinary} linear second order differential equation,
\bea
\label{Req}
\left ( \p_r^2 -k^2 - { f'(r) \over f(r) } \p_r \right ) \Phi _k (r)=0
\eea
Denoting the two linearly independent solutions to this equation by
$R^{(1)} _k (r)$ and $R^{(2)} _k (r)$, the general solution to (\ref{Req})
takes the form,
\bea
\Phi _k (r) = \phi _1 (k) R^{(1)} _k (r) + \phi _2 (k) R^{(2)} _k (r)
\eea
in terms of  two arbitrary complex functions $\phi_1(k)$ and $\phi _2 (k)$. 
The Wronskian of these two solutions is readily computed,
$\p_r R^{(1)} _k (r) R^{(2)} _k (r) - R^{(1)} _k (r) \p_r R^{(2)} _k (r) 
= w_0 f(r)$, were $w_0$ is found to be independent of $r$.

\subsection{A simple example }

A simple, but non-trivial, generalization of (\ref{M}) is obtained by taking $f(h)= h^{2 \ell-1}$
for real~$\ell$. The ordinary differential equation for the Fourier modes then takes
the form,
\bea
\label{R}
\left ( \p_r^2 - k^2 - {2\ell -1 \over r} \p_r \right ) R _k (r)=0
\eea
The linearly independent solutions to this equation may be chosen as follows,
\bea
R^{(1)}_k (r) & = & r^\ell I_\ell (kr)
\no \\
R^{(2)}_k (r) & = & r^\ell K_\ell (kr)
\eea
where $I_\ell$ and $K_\ell$ are the modified Bessel functions of order $\ell$
(which need not be an integer). For $\ell=1$, we recover the results of
\cite{D'Hoker:2008wc}. For $\ell$ integer, one may also derive this solution
via projection of a harmonic function in 3 dimensions, following the methods
of \cite{D'Hoker:2008wc}.

\section{Enlarging the integrable system}
\setcounter{equation}{0}

The reduction of the BPS equations in Type IIB supergravity \cite{D'Hoker:2007xy}, 
produced the  linear equation for $\p_w F$ of (\ref{F}), as well as a supplementary 
inhomogeneous linear equation for $\p_w \bar F$. Translated to the notation of the 
present paper, this {\sl supplementary equation} takes the form, 
\bea
\label{pbarF}
\p_w \bar F = \bar F \, \p_w \ln (1+f) + \kappa \, \sqrt{f}
\eea
where $f$ is given by equation IIB of (1.4), and $\kappa = \kappa (w)$ is an arbitrary 
holomorphic 1-form on $\Sigma$.
The integrability condition between (\ref{pbarF}) and (\ref{F}) is automatically satisfied. 

\sm

In this section, we shall show that such a supplementary equation exists for all $f$, 
and all $\kappa$. 
To derive it, we change variables to a more convenient field $G$, defined by 
$F = \sqrt{f} G$. In terms of $G$, (\ref{F}) takes the form, 
\bea
\label{G1}
\p_w G = \half \bar G \, \p_w \ln f(h)
\eea
and we postulate the following Ansatz for the supplementary equation, 
\bea
\label{G2}
\p_w \bar G = g_1 \, G \, \p_w h + g_2 \, \bar G \, \p_w h + g_3 \, \kappa
\eea
Here, the functions $g_1, g_2, g_3$ are complex and depend on $h$ only (but this 
dependence may involve $f$), and $\kappa$ is an arbitrary given holomorphic 
one-form on $\Sigma$. Simultaneous integrability of (\ref{G1}) and (\ref{G2})
is achieved by requiring that the $\p_{\bar w}$-derivative of (\ref{G1})
and the $\p_w$ derivative of the complex conjugate of (\ref{G2}) be equal,
producing the following equations,  (we shall denote 
derivatives with respect to $h$ by a prime, and use the abbreviation $\gamma = f'/f$),
\bea
\label{geq}
 \gamma^2 & = & 4 |g_1|^2 + 4 \bar g_2 '
\no \\
 \gamma' & = & 2 \bar g_1 ' + 2 g_2 \bar g_1 + \gamma \bar g_2
\no \\
0 & = & \bar g_1 g_3 \kappa \p_{\bar w} h + \bar g_3 ' \bar \kappa \p_w h
\eea
We shall show that this system always admits solutions for which the
functions $g_1,g_2,g_3$ are real, and henceforth specialize to this case.
The first two equations may be recast as follows,
\bea
\label{G3}
4 g_2 ' & = & (\gamma + 2 g_1) ( \gamma - 2 g_1)
\no \\
(\gamma - 2 g_1)' & = & g_2 (\gamma + 2 g_1)
\eea
from which we obtain the first integral, $4 g_2 ^2 - (\gamma - 2 g_1)^2 = -16\nu^2$,
where $\nu^2 $ is arbitrary real and independent of $h$. 
The solutions to this first integral may be 
parametrized as follows,
\bea
 g_2 & = & 2\nu \, \sh ( \phi)
\no \\
\gamma - 2 g_1 & = & 4 \nu \, \ch (\phi)
\eea
while the remaining equation in (\ref{G3}) gives $\gamma + 2 g_1 = 2 \phi'$.
The result is a single equation for $\phi$ in terms of $\gamma $, given by 
$ \phi ' =  \gamma - 2\nu \, \ch (\phi)$.
A change of the variable $\phi$ to $\psi = e^\phi /\sqrt{f}$ leads to an
equation of the Riccati type,
\bea
\psi ' = - {\nu \over  f} ( f^2 \psi ^2 + 1)
\eea
which may be reduced to a system of two linear equations by the 
customary change of variables, $\psi = \alpha / \beta$ with $\alpha$ and $\beta $ real, 
and we get
\bea
\alpha ' & = & - {\nu \over  f}  \,  \beta 
\no \\
\beta ' & = &  f  \nu \,  \alpha
\eea
The corresponding second order equations solely for $\alpha$ and $\beta$ are then,
\bea
\alpha '' + { f' \over f} \alpha ' + \nu ^2 \alpha & = & 0
\no \\
\beta '' - { f' \over f} \beta ' + \nu ^2 \beta & = & 0
\eea
which may be solved in terms of the functions $R^{(1,2)}_k(r)$ introduced in 
(\ref{Req}), upon analytic continuation of the equation under $k \to i \nu$.

\sm

It remains to solve the third equation of (\ref{geq}). We may assume that
$g_3 \kappa  \not=0$, since in the contrary case the third equation is 
manifestly satisfied. There are now two cases,
\bea
{\rm case ~ I} ~ & \hskip 0.5in & \kappa \p_{\bar w} h \pm \bar \kappa \p_w h \not =0
\no \\
{\rm case ~ II} && \kappa \p_{\bar w} h - \bar \kappa \p_w h  =0
\eea
(A third possibility, for which $\kappa \p_{\bar w} h + \bar \kappa \p_w h  =0$,
is equivalent to case II.) In case I, we must have $g_1g_3=g_3'=0$, which 
implies $g_1=0$ since $g_3 \not=0$. This case is easily solved and requires
that $f$ be given by an immediate generalization of the expression for the 
Type IIB case of (\ref{MIIB}), namely $f(h) = c_0 \tg (c_1h + c_2)^2$, with
$c_0, c_1,c_2$ constants, $c_0$ real, and $c_1,c_2$ either both real 
or both imaginary.
In case II, we must have $g_1g_3+g_3'=0$, and we may solve for $g_3$
as follows, $g_3 = g_0 \beta /\alpha$.

\newpage

\noindent
{\large \bf Acknowledgments}

\bigskip

We wish to acknowledge our collaborators  Michael Gutperle and Darya Krym 
on the supergravity solution papers, upon which the generalization derived in this 
note is based. We have also benefited from  conversations with Olivier Babelon
and Igor Krichever.

\end{document}